\documentclass[11pt,a4paper]{JHEP3}
\pdfoutput=1
\usepackage{graphicx}
\usepackage{bm}
\usepackage{mathrsfs}
\usepackage{amsmath}
\usepackage{amssymb}
\usepackage{amstext}
\usepackage{epsfig,latexsym}

\title{Massive Quantum Liquids from Holographic Angel's Trumpets}
\author{
Matthias C. Wapler\\
Center for Quantum Spacetime, Sogang University, Seoul, Korea \\
\\E-mail: \email{wapler@sogang.ac.kr}}
\preprint{}
\newcommand{\tlrho}{{\tilde{\rho}}}
\newcommand{\tlc}{{\tilde{c}}}
\newcommand{\tlm}{{\tilde{m}}}
\newcommand{\tlmu}{{\tilde{\mu}}}
\newcommand{\tlt}{{\tilde{t}}} 
\newcommand{\tx}{{\tilde{x}}}

\newcommand{\barho}{{\bar{\rho}}}
\newcommand{\bac}{{\bar{c}}}
\newcommand{\baT}{{\bar{T}}}
\newcommand{\KkappaK}{K}
\newcommand{\order}{{\mathcal{O}}}
\newcommand{\labell}[1]{\label{#1}}
\newcommand{\reef}[1]{(\ref{#1})}
\newcommand{\mref}[1]{\ref{#1}}
\newcommand{\mlabel}[1]{\label{#1}}
\date{\today}
\abstract{%
We explore the small-temperature regime in the deconfined phase of massive fundamental matter at finite baryon number density coupled to the 3+1 dimensional $\mathcal{N}=4$ SYM theory. In this setting, we can demonstrate a new type of non-trivial temperature-independent scaling solutions for the probe brane embeddings.

Focusing mostly on matter supported in 2+1 dimensions, the thermodynamics indicate that there is a quantum liquid with interesting density-dependent low-temperature physics. We also comment about 3+1 and 1+1 dimensional systems, where we further find for example a new thermodynamic instability.}

\begin{document}
\section{Introduction}
In recent years, the ADS/CFT correspondence \cite{adscft_Maldacena,adscft_gubser,adscft_witten} has become a powerful tool to study various properties of the strong coupling limit of conformal field theories, with applications to QCD and also more recently to some aspects of condensed matter physics. In principle, one has to differentiate between top-down setups that are constructed within string theory and imply consistency -- and bottom-up setups, where the gravitational duals are constructed from a phenomenological point of view. In this paper, we use the former approach as we would like to explore what happens to a particular consistent theory. 

In condensed matter applications, there has been particular interest in 2+1 dimensional systems that can be typically constructed using M2 branes \cite{pavel}, or as a defect in a 3+1 dimensional background using D3-D5 \cite{fancycon,conpaper} and D3-D7 \cite{fancycon,conpaper,rey} intersections -- and also as bottom-up setups in various contexts such as superconductivity. As our world is 3+1 dimensional, the defect setup may be more realistic, even though there are some problems with the consistency of the D3-D7 systems \cite{conpaper,fancycon}. 
Recently, there has also arisen significant interest in fermi-liquid-like aspects at low temperature compared to the density, for example in a 3+1 dimensional D3-D7 setup \cite{zerosound} or a 1+1 dimensional D3-D3 configuration \cite{luttinger}. This low-temperature limit will also be in the focus of this paper.

A common way how to introduce fundamental matter in ADS/CFT is in probe brane configurations, where one considers an $AdS$ blackhole background. For example one considers the well-known $AdS_5\times S^5$ (black hole) solution (above the deconfinement phase transition) from the decoupling limit of a stack of $N_c\gg 1$ D3 branes, that is dual to a (thermal) $SU(N_c)$ $\mathcal{N}=4$ supersymmetric Yang-Mills theory \cite{adscft_gubser}. Then one inserts $N_f \ll N_c$ intersecting Dp branes, giving $N_f$ families of (charged) fields in the fundamental representation of the $SU(N_c)$, living along the directions of the intersection \cite{winters,robfirst} -- such as the above-mentioned D3-Dp intersections.

In these setups, the mass $M_q$ of the fundamental matter is usually given by a scalar describing the size of the compact sphere of the probe brane geometry inside the $S^5$ \cite{winters}. Typically large masses or small temperatures correspond to a vanishing size of the compact sphere away from asymptotic infinity, i.e. to a ``narrow funnel'' that depends for example on the baryon number density $\rho$. 
However it was noted in \cite{fancytherm} 
that in 
$d +1$ dimensional probe brane setups
there is a coincidence in the scalings at finite $\rho$ and $M_q$: The relevant parameters are dimensionless with appropriate powers of the temperature and at small temperatures the dimensionless mass scales as
$\frac{M_q}{T} \sim \alpha \left(\frac{\rho}{T^{d}}\right)^{1/d}$, where the proportionality constant $\alpha$ depends  on the ``size of the funnel''.
Hence, the exponents of $T$ cancel and the size of the compact sphere remains finite and independent of temperature. 
This can be visualized in the usual picture by the extra tension or "stiffness" of the probe branes from the increasing $\frac{\rho}{T^d}$, as $T$ decreases, which will be illustrated in figure \ref{physembed}.
%
%
%
%
%
%
%
%
%
%
%
%

This scaling coincidence suggests that there may be some interesting low-temperature physics in this limit, and hence we will explore the details of the embeddings and their thermodynamics and related physical properties. We will mostly consider the 2+1 dimensional defect setup, but eventually we will also briefly look at the results in 3+1 and 1+1 dimensions, and point out some important differences.

This paper is organized as follows: First, we review the string theory setup and the necessary ADS/CFT dictionary of the D3-D5 defect in section \ref{setup}. Then, we will demonstrate the scaling solution for the probe-brane embedding that is central to this paper in section \ref{limit} where we also comment on the relevant thermodynamic variables to consider in the massive case. In section \ref{thermo}, we first obtain the leading terms of the free energy, then discuss the thermodynamics/statistical mechanics in 2+1 dimensions and finally study the subleading terms that give rise to the heat capacity. Throughout this section, we will see how the system interpolates between a mass-dominated limit and a density-dominated limit. Finally, we will briefly discuss the generalized 3+1 and 1+1 dimensional systems in section \ref{gencas};
and then we will discuss the results and conclude in section \ref{conclu}.
\section{Setup}\label{setup}
We start with the supergravity background of a planar black hole in $AdS_5$, 
\begin{eqnarray}
ds^2 & =& \frac{r^2}{L^2} \left( -(1-r_0^4/r^4)dt^2 +d\vec{x}_3^2\right)
+ \frac{L^2}{r^2} \left( \frac{dr^2}{1-r_0^4/r^4} +r^2 d\Omega_5^2\right) \
, \ \ \  C^{(4)}_{txyz}=-\frac{r^4}{L^4} \labell{D3geom} \ .
\end{eqnarray}
This
corresponds to the decoupling limit of $N_c$ black D3-branes
dual to ${\cal N}=4$ $SU(N_c)$ super-Yang-Mills theory at finite temperature $T$, living along the flat directions of the AdS, \cite{adscft_witten}.
The  temperature $T$ is given by the Hawking temperature $T = \frac{r_0}{\pi L^2}$ and the Yang-Mills coupling by $g_{ym}^2 = 4 \pi g_s$. Since the curvature $L$ is
given in terms of the string coupling $g_s$ and 
string length  $l_s$ as $L^4 = 4\pi\, g_s N_c \, l_s^4$, the 't Hooft
coupling $\lambda=g_{ym}^2 N_c$ can be written as $\lambda=\frac{L^4}{l_s^4}$. Hence the ``supergravity limit'' $L \gg l_s$ in which the type IIB supergravity action and the solution \reef{D3geom} are valid  corresponds to strong coupling $\lambda \gg 1$.

In practice, however, we will use coordinates that are made dimensionless with factors of $\frac{L^2}{r_0} =\frac{1}{\pi T}$, denoted by $\tilde{(\cdot)}$, such as $\tlt: = \frac{r_0 \, t}{L^2} =  t\, (\pi T)$, and the inverse dimensionless radius $u := \frac{r_0}{r}$, giving us 
\begin{equation}\label{branemetric}
ds^2 \, = \, L^2 u^2\left(-(1-u^4)d\tlt^2 + d\vec{\tx}_3^2 + \frac{du^2}{1-u^4} + d\Omega_5^2 \right) \ .
\end{equation}

Because in this setup all the fields transform in the adjoint representation of the $SU(N_c)$ its use in QCD or condensed matter physics is very much limited as one would like to consider also matter that is charged under this symmetry, i.e. that transforms in the fundamental representation.
To introduce the fundamental matter one then creates an intersection of ``probe'' Dp branes with the D3 branes, such that at the string theory side there are fields at the massless level of field theory at the intersection. From the point of view of the probe branes, they correspond to endpoints of D3-Dp strings, and in the gravity side they correspond to fundamental fields in the (defect) field theory.

Here, we use the well- known D3-D5 defect setup (see e.g. \cite{lisa,hirosi,johannadef}):
\begin{equation}
\begin{array}{rccccc|c|cccccl}
  & & 0 & 1 & 2 & 3 & 4& 5 & 6 & 7 & 8 & 9 &\\
  & & t & x & y & z & r&   &   &   &   &  \theta &\\
\mathrm{background\,:}& D3 & \times & \times & \times & \times & & &  & & & & \\
\mathrm{probe\,:}& D5 & \times & \times & \times &  & \times  & \times & \times & &  & &  \ \ \ .
\end{array} 
\labell{array}
\end{equation}
The dual field
theory is now the SYM gauge theory coupled to $N_f$ fundamental
hypermultiplets, which are confined to a (2+1)-dimensional defect.
This construction is still supersymmetric, but the supersymmetry has been reduced from ${\cal N}=4$ to
${\cal N}=2$ by the introduction of the defect. In the limit $N_f
\ll N_c$, the D5-branes may be treated as probes in the
supergravity background, i.e. we may ignore their gravitational
back-reaction.

Here we want to turn on only the overall $U(1)$ factor of the world-volume gauge field, so the the probe branes are governed by the DBI action
\begin{equation}\label{braneaction}
S \, = \, - T_5 N_f \int_{D5} \sqrt{- det(P[G] + 2\pi l_s^2 F )}   \ ,
\end{equation}
evaluated in the D3 background \reef{D3geom}.
We assume also translational invariance along the flat directions and rotational invariance on the sphere. Hence, the pullback in \reef{braneaction} gives us one scalar field corresponding to the position in the $z$ direction, which was extensively studied in \cite{conpaper,fancycon}, and another scalar which describes the size of the compact sphere and corresponds to turning on the mass of the fundamental matter, studied in \cite{fancytherm,fancycon} and more extensively in the similar D3-D7 system in \cite{long,johanna,ingo,robdens,robchem}. Parametrizing the $S^5$ as 
$d\Omega_5^2 = d\theta^2 + \sin^2 \theta\, d\Omega_2^2 + \cos^2 \theta\, d\Omega_2^2$ and putting  the brane on the first $S^2$ of the $S^5$,
the induced metric on the probe branes is given by
\begin{equation}\labell{branemet}
ds^2 \, = \, \frac{L^2}{u^2} \left( -(1- u^4)d\tlt^2  + d\vec{\tx}_2^2  + \left(\frac{1}{1-u^4}+ \frac{u^2 \Psi'(u)^2}{1-\Psi(u)^2}  \right) du^2  \, + \, u^2 (1-\Psi(u)^2) d\Omega_2^2\right) \ ,
\end{equation} 
where we defined the scalar as $\Psi(u) = \sin \, \theta(u)$.

We also choose to turn on the world-volume $U(1)$ gauge field which is dual to the $U(1)$ current operator of the $U(N_f)$ that gives rise to the ``flavor symmetry''. In particular, the flux of the electric field
\begin{equation}
F \ = \ \partial_u A_{t}(u) \, du \wedge dt \ ,
\end{equation}
corresponds in the field theory side to the baryon number density (see e.g. \cite{bigrev})
\begin{equation}\label{rhodic}
\rho  \ = \ \big<\!\ J^t \big> \ = \ - \frac{1}{N_c}\frac{\delta S}{\delta A_{t}^{bdy.}} \ = \  4\pi  N_f \frac{L^2 T_5}{N_c r_0} \lim_{u\rightarrow 0} \partial_u A_t(u) \ .
\end{equation}

In this parametrization, the DBI action becomes
\begin{equation}\labell{backact}
S \ = \ 4 \pi L^2 T_5 N_f \int\! d\sigma^4 \sqrt{-\det P[G]_4}\sqrt{1-( 2\pi l_s^2 \partial_u {A}_t(u))^2P[G]^{tt}P[G]^{uu} }\, (1-\Psi^2) 
\end{equation}
where the integral is taken over the $AdS_4$ part of the world-volume.
Using this action, it is straightforward to obtain the solution for the gauge field
\begin{equation}\labell{backgd}
\ \partial_u {A}_t(u) \ = \ \sqrt{\lambda} T \frac{\tlrho\sqrt{1- \Psi^2(u) + u^2 h(u) \Psi'(u)^2}}{\sqrt{1-\Psi(u)^2}\sqrt{  \tlrho^2 u^4 + (1-\Psi(u)^2)^2}}  \nonumber,
\end{equation} 
where $\tlrho :=   \frac{\rho}{ N_f T^2}$.
The equation of motion for $\Psi(u)$ becomes
\begin{eqnarray}\labell{psieom}
\!\!\!\!\!\!\!\!\!\!\frac{2(1-\Psi^2)^3 +u^2 (1-u^4)\big(\tlrho^2 u^4+(1-\Psi^2)^2 \big)\Psi'^2}{u^4 (1\!-\!\Psi^2)\sqrt{(1\!-\!\Psi^2)\big(1\!-\!\Psi^2+(u^2\!-u^6)\Psi'^2 \big)\big(1+\tlrho^2 u^4 + \Psi^2(\Psi^2\!-2)\big)}} \nonumber \\
\!\!\!\!\!\!\!\!\!\! \!\!\!\!\!\!\!\!\!\! = \
\partial_u  \left(\Psi' \frac{1-u^4}{u^2}\sqrt{\frac{\tlrho^2 u^4 + (1-\Psi^2)^2}{(1-\Psi^2)(1-\Psi^2+(u^2-u^6)\Psi'^2)}} \right)~~~~~~~~~~~~~
 \ , 
\end{eqnarray}
and on the horizon $u=1$, the equations reduce to 
\begin{equation}\labell{psibdy}
\lim_{u\rightarrow 1^-} \partial_u \Psi\, = \, \frac{1}{2}\frac{\Psi_0(1-\Psi_0^2)^2}{\tlrho^2 + (1-\Psi_0^2)^2}  \ , \ \mathrm{where} \ \Psi_0 \, = \, \lim_{u\rightarrow 1^-} \Psi\ .
\end{equation}

The asymptotic solution at $u \rightarrow 0$ is
\begin{equation}\labell{psiasym}
\Psi \, \sim \, \tlm \, u \, + \, \tlc \, u^2 \ ,
\end{equation}
where $\tlm$ and $\tlc$ are dimensionless free parameters that are determined by the value on the horizon $\Psi_0$ and \reef{psibdy} that acts as a second boundary condition on the horizon. Following arguments of the T-dual $(3+1)$ dimensional D3-D7 setup \cite{long,johanna,ingo,robdens,robchem}, the quark mass $M_q$ and dual condensate $C$ are given by 
\begin{equation}
M_q \, = \, \frac{r_0 \,\tlm}{2 \pi l_s^2} \, = \,  \frac{\sqrt{\lambda}}{2} T \tlm\ \ \  \mathrm{and}  \ \ \  C\, = \,  T^2 N_f N_c\tlc\ .
\end{equation}
One can understand this identification of the mass from the separation between the D3 and D5 branes in flat space, such that $M_q$ is the mass of a stretched D3-D5 string and the condensate is just the thermodynamic dual of the mass.

At vanishing density, there is a critical temperature-mass ratio below which the probe branes do not extend down to the horizon \cite{johannaphase,long}. At finite densities, and unless one turns on the scalar in the $z$ direction considered in \cite{fancycon,fancytherm}, the brane always extends down to the horizon even though a phase transition may still be observed at small densities \cite{fancytherm}. In the limit considered in this paper, however, this phase transition is of no concern.

\begin{figure}
\includegraphics[width = 0.99\textwidth]{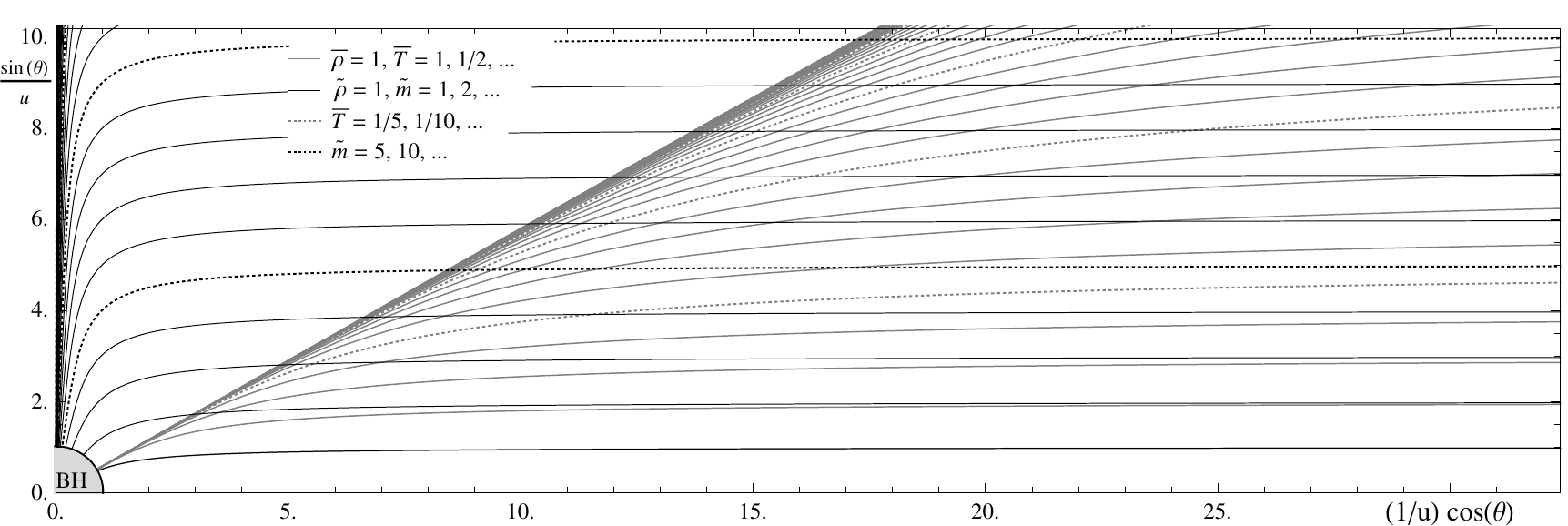}
\caption{The full ``physical embedding'' of the probe brane, represented by the ``radial'' direction of the bulk spacetime $\frac{1}{u}\cos\, \theta$ and the ``separation'' $\frac{1}{u}\sin\, \theta$ for fixed $\tlrho = 1$ (at fixed temperature and varying mass $\tlm$, black) and fixed $\barho = \frac{\tlrho}{\tlm^2} = 1$ (at fixed mass and varying temperature $\baT = \frac{1}{\tlm}$, gray). The corresponding curves of both cases should meet at asymptotic infinity, $\frac{1}{u}\cos\theta \rightarrow \infty$}\label{physembed}
\end{figure}

In general, equation \reef{psieom} has no analytical solution and \reef{psibdy}  implies that one has to start integrating the equation from the horizon to obtain the mass and the related condensate at the asymptotic boundary, rather than setting either of them first. We show the full numerical solutions for $\Psi(u)$ in figure \ref{physembed}, where we illustrate the ``physical'' embedding $\left(\frac{1}{u} \cos\theta ,\frac{1}{u} \sin\theta\right)$ of the probe brane for both fixed $\rho/T^2$ and fixed $\rho/M_q^2$ to support the physical interpretation outlined already in the introduction. 

In the following section, however, we will demonstrate a particular case in which the equation of motion for $\Psi$ simplifies dramatically.
\section{Taking the limit}\label{limit}
In this section, we obtain the scaling solution in a ``blow-up region'' near asymptotic infinity $u\rightarrow 0$ at small temperatures and finite density and mass, i.e. $\tlrho \gg 1$, $\tlm \gg 1$.

%
Before we do so, we need to first consider an approximate solution at finite values of $u$ in order to match the asymptotic solution to the black hole horizon.
Hence, let us consider a linear expansion around an approximately finite size funnel, $\Psi(u) \ = \ \Psi_0 + \psi(u)$, which reduces \reef{psieom} to
\begin{equation}
2\Psi_0(1- \Psi_0^2)\ =\ \tlrho^2 u^6 \partial_u \left((1-u^4) \partial_u \psi(u)\right) \ +\ \order(\psi^2)\ .
\end{equation}
Enforcing the boundary condition \reef{psibdy} and $\psi = 0$ at the horizon gives the solution
\begin{equation}\labell{psilogsol}
\psi(u) \, = \, \frac{-\Psi_0(1-\Psi_0^2)}{10\, \tlrho^2 u^4}\left(1-u^4 \, - 2 u^4\tan^{\! -1}\frac{1-u}{1+u}  +  u^4 \ln\frac{(1+u^2)(1+u)^2}{8 \, u^4}  \right) \,+  \order(\tlrho^{-4}) \ ,
\end{equation}
which is negative over $u\in ]0,1[$ and becomes finite at small $u\sim \order(1/\sqrt{\tlrho})$ as it should. 

Now we wish to find some overlap of this solution with a solution in the blow-up region at small $u$ that we can match around $\frac{1}{\sqrt{\tlrho}} \ll u \ll 1$. To do so, we define the ``blow-up'' coordinate $\xi:=u \sqrt{\tlrho}$, which maps the horizon to large $\xi = \sqrt{\tlrho} \sim T^{-1} \gg 1$, and the scale over which most of the change in $\Psi$ takes place to $\xi \in ]0,\order(1)]$. Then, we consider all of $\xi\in \mathbb{R^+}$ and expand for large $\tlrho$. 
The leading term (up to order $\tlrho^{-2}$) is written in the most compact form as
\begin{eqnarray}
0 & = &  \partial_\xi\left(\sqrt{1-\Psi(\xi)^2}\sqrt{\xi^4 + (1-\Psi(\xi)^2)^2} \partial_\xi \left(\frac{\Psi(\xi)}{\xi}\right) \right) \\ \nonumber & & + \ 
\frac{\partial_\xi\Psi(\xi)}{\Psi(\xi)} \left(\xi \partial_\xi \left(\frac{\Psi(\xi)}{\xi}\right)\right)^2
\partial_{\Psi(\xi)}\left(\sqrt{1-\Psi(\xi)^2}\sqrt{\xi^4 + (1-\Psi(\xi)^2)^2} \right) \ ,
\end{eqnarray}
where the erratic solution $\Psi = m \, \xi$ has to be ignored.
It is straightforward to show that the expansion for large $\xi$ is indeed consistent with $\Psi  = \Psi_0 - \frac{\Psi_0 (1-\Psi_0^2)^2}{10 \xi^4}$, i.e. with the small-$u$ limit of \reef{psilogsol}. In practice for the numerics, we have to choose $\xi^{-4}$ as a coordinate and implement this matching as a boundary condition as $\xi^{-4} \rightarrow 0$. 

Because we want to consider fixed mass, but vary the temperature and density, the suitable physical quantities to consider at finite mass are 
\begin{eqnarray}
\baT & =& \frac{1}{\tlm} \ = \ \sqrt{\lambda} \frac{T}{2 M_q} \ , \\
\barho & = & \frac{\tlrho}{\tlm^2} \, =\,  \frac{ \rho}{ N_f}\frac{\lambda}{4 M_q^2} \ \mathrm{and} \\
\bac & =& \frac{\tlc}{\tlm^2} \ =\   \frac{C}{ N_c N_f}\frac{\lambda}{4 M_q^2} \ ,
\end{eqnarray}
where the numerical coefficients are chosen to give
us parameters that are straightforwardly related to the embedding.
Using \reef{rhodic}, \reef{psiasym} and the definition of $\xi$, they are now related to the scalar by 
\begin{equation}
\barho \, = \, (\partial_\xi \Psi |_{\xi\rightarrow 0})^{-2} \, ,  \ \ \ \bac\, =\, \barho \frac{1}{2}\partial_\xi^2 \Psi |_{\xi\rightarrow 0} \ \ \mathrm{and} \ \ \ \baT\, =\, \sqrt{\barho/\tlrho} \ .
\end{equation}
It turns out that there seems to be an exact relation $\partial_\xi^2 \Psi |_{\xi\rightarrow 0} = - 2 \Psi_0$, such that $\bac =- \barho\, \Psi_0$.

\begin{figure}
\includegraphics[width = 0.49\textwidth]{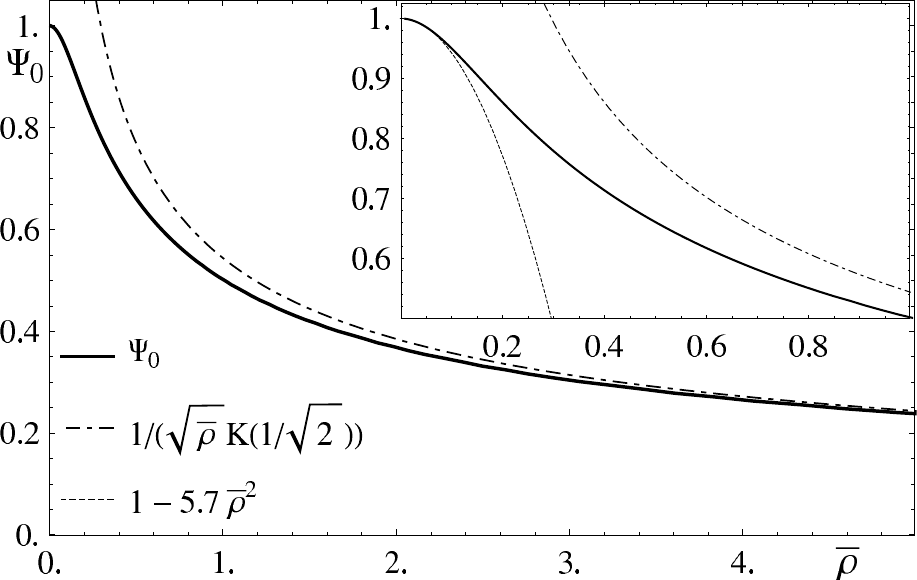} 
\hspace{0.02\textwidth} 
\includegraphics[width = 0.49\textwidth]{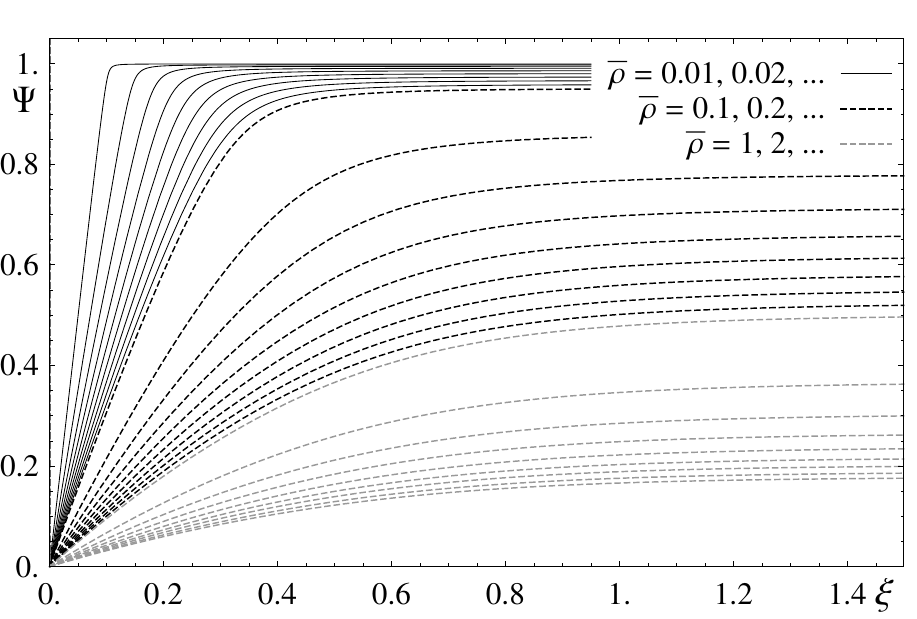} 
\caption{Left: $\Psi_0(\barho)$ with the approximations at small and large $\barho$. Right: $\Psi(\xi)$ for various values of $\barho$.}\label{emplot}
\end{figure}

For small $\Psi_0$, we can expand the equations of motion to linear order,
\begin{equation}
0 \ = \ 2\Psi \, - \, 2 \xi \partial_\xi \Psi \, + \, \xi^2(1+\xi^4)\partial_\xi^2\Psi \ .
\end{equation}
This has an analytic  solution in terms of the elliptic integral of the first kind, $\mathcal{F}(\phi|k) = \int_0^\phi \frac{d\varphi}{\sqrt{1-k^2\sin^2\varphi}}$ and the complete elliptic integral $\mathcal{K}(k)=\mathcal{F}\!\left(\frac{\pi}{2}\big|k\right)$
\begin{equation}\labell{psiK}
\Psi \ = \ \Psi_0 \xi \left( e^{i\pi/4} \mathcal{F} \left(i \sinh^{-1}(e^{i\pi/4} \xi) \Big| i\right) \ + \ \mathcal{K}(1/\sqrt{2})\right) \ ,
\end{equation}
which is itself not particularly interesting because it just interpolates between the linear behavior at small $\xi$ and the constant $\Psi_0$ at large $\xi$; and since it is the linear expansion in $\Psi_0$ it tells us that at large densities the relevant scales in $\xi$ do not depend on the density. The profile can be seen in the curves with large $\barho$ on the right in fig. \ref{emplot}.
The important point is that because $\forall k: \, \mathcal{F}(x|k) \sim x + \order(x^2)$, we can straightforwardly read off the first and second derivatives at $\xi \rightarrow 0$ to find that 
\begin{equation}
\Psi_0 \, \sim\, \frac{1}{\sqrt{\barho} \, \mathcal{K}(1/\sqrt{2})}\ \ \mathrm{and} \ \ \  \bac \, \sim\, -\Psi_0\, \barho\, \sim \, \frac{\sqrt{\barho}}{ \mathcal{K}(1/\sqrt{2})} \ .
\end{equation}

In the regime of $\Psi -1 \ll 1$, however, one cannot find such a simple expression. By studying the equations of motion, one can only argue that the transition from the large-$\xi$ solution to the asymptotic solution occurs at $\xi\sim\order(1-\Psi_0)^{(1/4)}$, such that $1-\Psi \propto \barho^2$. 
The corresponding value of the condensate is $\bac = - \barho + \order (\barho^3)$ for small values of $\barho$.

\begin{figure}
\includegraphics[width = 0.49\textwidth]{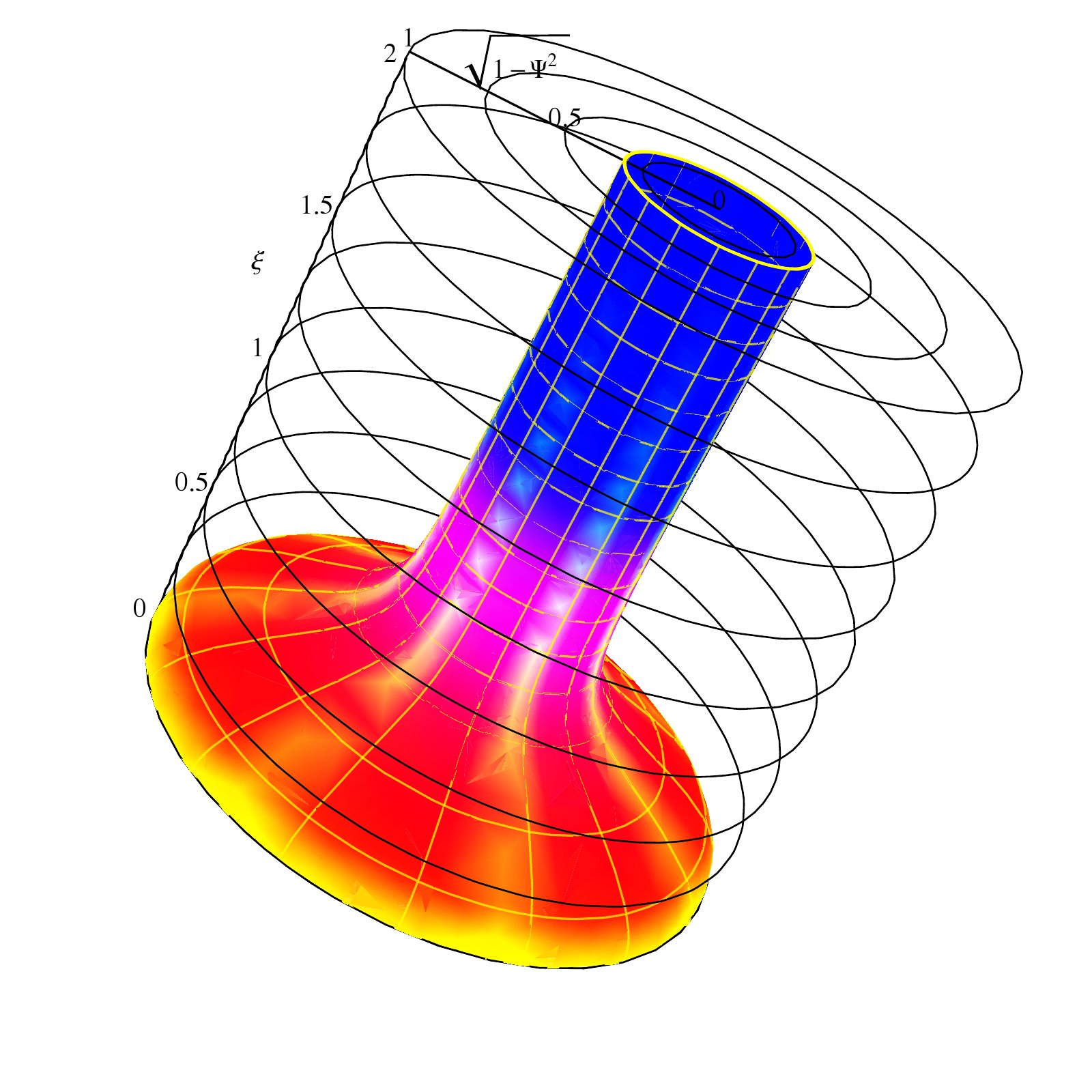} 
\hspace{0.02\textwidth} 
\includegraphics[width = 0.49\textwidth]{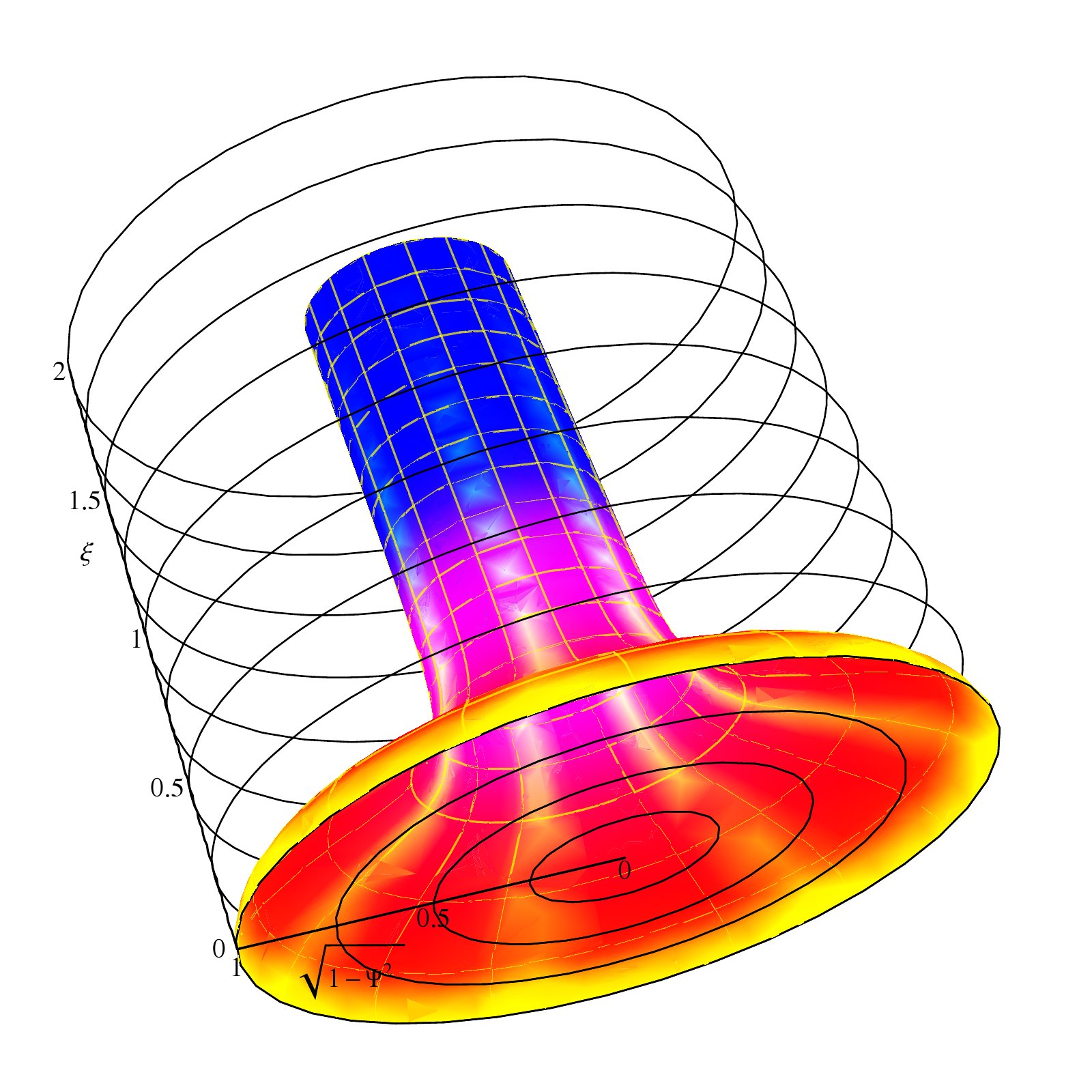} 
\caption{Illustration
of the size of the temperature-independent size $\cos\,\theta = \sqrt{1-\Psi^2}$ of the compact $S^2$ (only $S^1$ shown here) in the asymptotic blow-up region as a function of the inverse radial coordinate $\xi$, for $\Psi_0 = 0.95$. The color scheme is an artist's impression to illustrate the similarity with the flowers of Angel's Trumpets ({\it Brugmansia}).}\label{fancyplot}
\end{figure}

The full numerical dependence $\Psi_0(\barho)$ and these limiting cases are shown in figure \mref{emplot}, where we also show the corresponding embeddings $\Psi_{\barho}(\xi)$.
We further illustrate\footnote{The color scheme is an artist's impression to illustrate the similarity with the flowers of Angel's Trumpets ({\it Brugmansia}).}
 the ``size'' of the sphere, $\cos \theta = \sqrt{1-\Psi^2}$ in figure \ref{fancyplot}.

\section{Thermodynamics}\label{thermo}
In order to study the thermodynamic quantities, we follow the standard procedure of obtaining the free energy at fixed quark mass as a function of density and temperature, $F=F_{M_q}(\rho,T)$, and then computing its derivatives.  
The extrinsic thermodynamic quantities of the defect should be considered as a contribution to the overall 3+1 dimensional system.
In \cite{fancytherm}, it was argued, however,
that
they can be discussed independently, since the asymptotic properties of the bulk do not depend on the details of the defect.
As usual, we consider all extrinsic quantities densitized and will comment on obtaining the pressure separately.
\subsection{Free Energy}
In \cite{fancytherm}, we computed the free energy density for this system via the standard procedure from the Euclidean action, $I_e$, \cite{long} using 
\begin{equation}\mlabel{freeendef}
F \ = \ T \tilde{I}_e \  ,
\ \ \ \tilde{I}_e \ = \  \int_{u_{min}}^{u_{max}} \mathcal{L}_e \ + \ I_{bdy} \ + \ \rho A_t\ ,
\end{equation}
where the boundary terms
\begin{equation}\label{ibdy}
I_{bdy.} \ = \ - \frac{1}{3} \sqrt{\gamma} + \frac{1}{2}\Psi^2\sqrt{\gamma}
\end{equation}
are dictated to us by consistency \cite{skenderis}. The Legendre transformation needs to be done in order to obtain $\rho$ as a thermodynamic variable, rather than the chemical potential (see e.g. \cite{robdens,fancytherm} for a more detailed discussion).
At large temperatures, i.e. $\tlm,\tlrho \ll 1$, this evaluates to
\begin{equation}
\bar{F} \ := \ F\frac{\lambda}{ 4N_c N_f M_q^3} \ = \ -\frac{1}{3} \baT^3 \ ,
\end{equation}
with the corresponding entropy $\bar{S} = \baT^2$ and heat capacity $\bar{c}_V = 2 \baT^2$. Note that the unusual factor $\lambda$ arises because of the definition of $\baT$.

To compute the value of the free energy in our case, we can again use $\xi=u \sqrt{\tlrho}$ and expand for large $\tlrho$ to obtain
\begin{equation}\labell{keyeq}
\bar{F} \ = \ \barho^{3/2}\left( \int \frac{d\,\xi}{\xi^4}\sqrt{\xi^4+(1-\Psi^2)^2}\sqrt{1+\frac{\xi^2\Psi'^2}{1-\Psi^2}} \ - \ \left.\frac{2-3\Psi^2}{6 \xi^3}\right|_{bdy.}\right) \ + \ \order(\tlrho^{-1/2}) \ .
\end{equation}
This has to be evaluated over $\xi \in ]0,\sqrt{\tlrho}]$, and we will try to split it into an integral over $\xi \in ]0,\infty [$ and $\xi \in ]\sqrt{\tlrho},\infty [$, for the latter of which we can consider a simple expansion of the integrand.
At large $\xi$, the integral becomes just $\int d\,\xi\,(\xi^{-2} \, + \, \order(\xi^{-6}))$. 
To see that
this is consistent at $\xi\sim\order(\sqrt{\tlrho})$ with the full expression for the integral found in \cite{fancycon},  we substitute the approximate the solution \reef{psilogsol} into the full expression and expand for large $\tlrho$ -- giving us $\int d u\,(\frac{\tlrho}{u^2} \, + \, \order(\tlrho^{-1}))$. 
In contrast to the procedure in \cite{fancytherm} and the ``main'' part of the integral $\xi \in ]0,\infty [$, we did not add the boundary term to this part of the integral. This is because it vanishes in the full solution on the horizon and in our approximate solution as $\xi \rightarrow \infty$, but not sufficiently fast near $\xi \sim \sqrt{\tlrho}$.

Putting both parts of the integral together, we find the result
\begin{equation}\labell{fformula}
\bar{F} \ = \ \barho^{3/2} \Phi(\barho) \, - \, \baT \barho \, + \, \order(\baT^4/\sqrt{ \barho}) \ ,
\end{equation}
where 
the term $\baT \barho$ comes from the integral over $\xi \in ]\sqrt{\tlrho},\infty [$
and
the integral
\begin{equation}\label{keyint}
\Phi(\barho) \ = \ \lim_{\varepsilon \rightarrow 0}\int_{\varepsilon}^{\infty}\frac{d\, \xi}{2 \xi^4}\left( 2 \sqrt{\xi^4+(1-\Psi^2)^2}\sqrt{1+\frac{\xi^2\Psi'^2}{1-\Psi^2}} -2 +3\Psi^2-2\xi\Psi\Psi'\right) \
\end{equation} 
is a function of $\barho$ only. An important observation is that the subleading term is of higher order in temperature than the $T^3$ behavior \ref{keyint} in the high-temperature regime.

At $\Psi = 0$, i.e. at  $\barho \rightarrow \infty$, the integral simplifies to $\int_0^\infty d\xi \frac{\sqrt{1+\xi^4} -1}{\xi^4} \, = \, \frac{2}{3}\mathcal{K}(1/\sqrt{2})$. We can also expand the integral to order $\Psi^2$ using the approximate solution for $\Psi$, equation \reef{psiK}, but the resulting integral can only be evaluated numerically and yields 
\begin{equation}
\Phi(\barho) \  = \  \frac{2}{3}\mathcal{K}(1/\sqrt{2})+ \frac{0.2697}{\barho} + \order(\barho^{-4}) \ .
\end{equation}

In the opposite limit at $\barho \ll 1$,
we approximate the solution by two integrals, below and above the ``singular'' point around $\sqrt{\barho}$, which is the ``kink'' in the profiles shown in figure \ref{emplot}.
The integral from $0$ to $\sqrt{\barho}$ with $\Psi \sim \xi/\sqrt{\barho} $ gives a contribution $ \frac{1}{6 \barho^{3/2}} + \order(1)$ and the integral from $1/\sqrt{\barho}$ to $\infty$ gives $ \frac{-1}{6 \barho^{3/2}}+\frac{1}{\sqrt{\barho}}$ to leading order in $1-\Psi_0$. Hence the leading behavior at small $\barho$ is 
$\Phi\sim \frac{1}{\sqrt{\barho}}$ or $\bar{F} \sim \barho(1-\baT)$. The next order in $\Phi(\rho)$ is numerically approximately $1.95 \barho^{3/2}$, i.e. $1.95 \barho^{3}$ in $\bar{F}$.

\begin{figure}
\includegraphics[width = 0.49\textwidth]{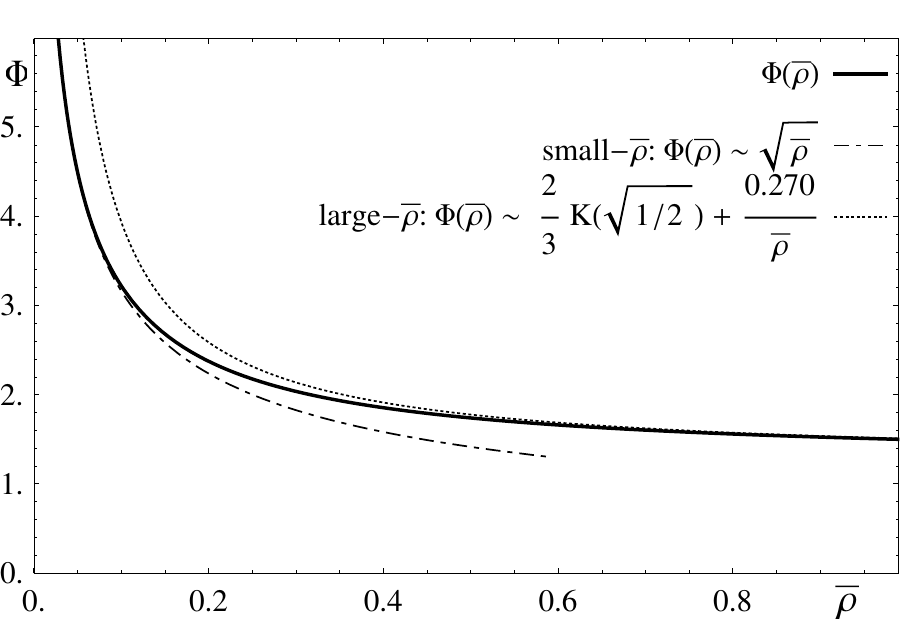}\hspace{0.02\textwidth}\includegraphics[width = 0.49\textwidth]{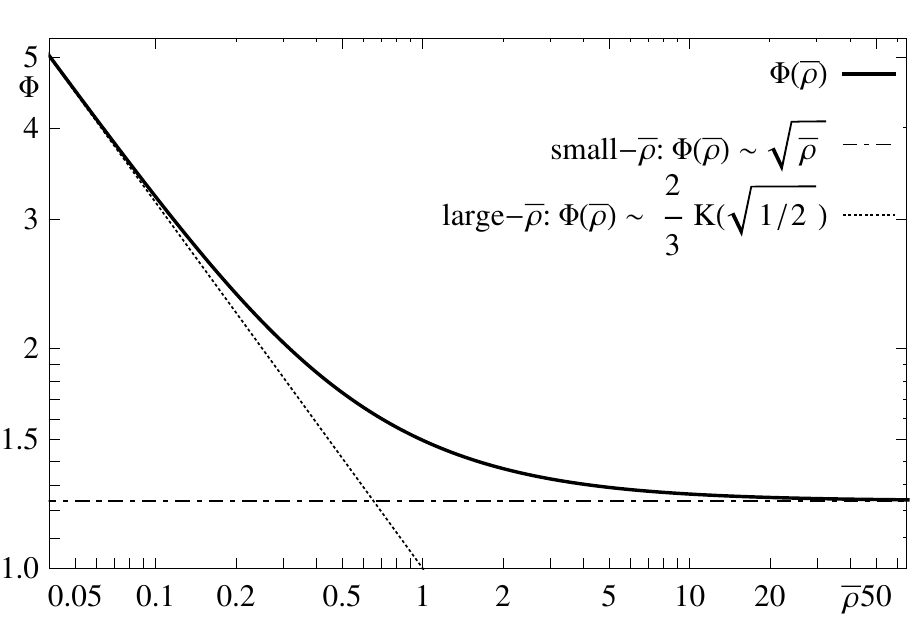}
\caption{$\Phi(\barho)$ as a linear plot (left) and log-log plot (right), illustrating the asymptotic limits.}\label{fofrhoplot}
\end{figure}

We show the full solution for $\Phi (\barho)$ together with the approximations in figure \ref{fofrhoplot}.

Reinstating the dimensionful coefficients, the free energy density becomes
\begin{equation}
F \, = \,  \frac{\sqrt{\lambda}}{2}N_c \rho \left( \sqrt{\frac{\rho}{N_f}}\Phi(\barho)\, - \, T\right)\, \sim \, \left\{\begin{array}{ll} 
N_crho\,  M_q - \frac{\sqrt{\lambda}}{2}N_c rho\,  T & \ \ : \frac{\rho}{ N_f} \ll \left(\frac{2M_q \!}{\sqrt{\lambda}}\right)^{\! 2} \\
 \frac{1}{3}\mathcal{K}(\frac{1}{\sqrt{2}})\, N_c\sqrt{\frac{\lambda}{ N_f}}\rho^{3/2}- \frac{\sqrt{\lambda}}{2}N_c rho\,  T & \ \ :
\frac{\rho}{ N_f} \gg \left(\frac{2M_q \!}{\sqrt{\lambda}}\right)^{\! 2}  \ .
\end{array}\right.
\end{equation}
\subsection{Thermodynamic quantities}\label{thermquan}
The entropy and internal energy densities are given by $S = -\left.\frac{\partial\, F}{\partial\, T}\right|_{\rho }$ and $U = F +TS$, which evaluates in our case to
\begin{eqnarray}
\!\!\!\!\!\!\!\! S & = & \frac{\sqrt{\lambda}}{2}N_c \rho \, + \, \order(\baT^3/\sqrt{ \barho}) \ \ \mathrm{and} \\
\!\!\!\!\!\!\!\! U & = &      \sqrt{\frac{\lambda}{N_f}}\frac{N_c}{2}\rho^{3/2}\Phi(\barho)\, +   \order(\baT^4/\sqrt{ \barho})\,  \sim   \left\{\begin{array}{ll} 
N_crho\,  M_q  & \ \ : \frac{\rho}{ N_f} \ll \left(\frac{2M_q \!}{\sqrt{\lambda}}\right)^{\! 2} \\
 \frac{1}{3}\mathcal{K}(\frac{1}{\sqrt{2}})\, N_c\sqrt{\frac{\lambda}{ N_f}}\rho^{3/2} & \ \ :
\frac{\rho}{ N_f} \gg \left(\frac{2M_q \!}{\sqrt{\lambda}}\right)^{\! 2}  \ .
\end{array}\right. ~~{ }
\end{eqnarray}
Correspondingly the specific heat capacity $c_V = \left.\frac{\partial \, U}{\partial\, T}\right|_{\rho }$ is 
\begin{equation}
c_v \ = \ 0\, + \,  \order(\baT^3/\sqrt{ \barho}) \ .
\end{equation}
We can also obtain the chemical potential $\mu = \left.\frac{\partial \, F}{\partial\, \rho}\right|_{T }$ , 
\begin{equation}
\mu \, = \, 
 \sqrt{\frac{\lambda}{N_f}}\frac{N_c}{2}\,
\partial_{\rho} 
\left(\rho^{3/2} \Phi(\barho)\right) 
\, - \,\frac{\sqrt{\lambda}}{2} T 
\, \sim \,
\left\{\begin{array}{ll} 
N_c M_q - \frac{\sqrt{\lambda}}{2}N_c  T & \ \ : \frac{\rho}{ N_f} \ll \left(\frac{2M_q \!}{\sqrt{\lambda}}\right)^{\! 2} \\
 \frac{1}{2}\mathcal{K}(\frac{1}{\sqrt{2}})\, N_c\sqrt{\frac{\lambda\, \rho}{ N_f}}- \frac{\sqrt{\lambda}}{2}N_c  T & \ \ :
\frac{\rho}{ N_f} \gg \left(\frac{2M_q \!}{\sqrt{\lambda}}\right)^{\! 2}  \ .
\end{array}\right.
%
\end{equation}
and the ``inverse density'' of states $\varepsilon^{-1} = \left.\frac{\partial^2 \, F}{\partial\, \rho^2}\right|_{T }$,
\begin{equation}
\epsilon^{-1} \, = \, 
\sqrt{\frac{\lambda}{N_f}}\frac{N_c}{2}\,
\partial_{\rho}^2 
\left(\rho^{3/2} \Phi(\barho)\right) 
 \, \sim \, \left\{\begin{array}{ll} 
0 + 0.73 \frac{N_c \lambda^2 \rho}{N_f^2 M_q^2}& \ \ : \frac{\rho}{ N_f} \ll \left(\frac{2M_q \!}{\sqrt{\lambda}}\right)^{\! 2} \\
 \frac{1}{4}\mathcal{K}(\frac{1}{\sqrt{2}})\, N_c\sqrt{\frac{\lambda }{\rho N_f}} & \ \ :
\frac{\rho}{ N_f} \gg \left(\frac{2M_q \!}{\sqrt{\lambda}}\right)^{\! 2}  \ .
\end{array}\right.
%
%
%
%
%
\end{equation}
where one has to keep in mind that $\barho = \frac{ \rho}{ N_f}\frac{\lambda}{4 M_q^2}$ and the subleading term in these expressions is of order $\baT^3$.

The density scaling of the chemical potential in the large-$\rho$ limit is in principle characteristic for a 2 dimensional Fermi liquid.
Interpreting this chemical potential however as a Fermi-sea -type chemical potential may be misleading, as the results of \cite{fancycon} suggest that the finite density gives rise to some finite length scale and to a quasiparticle spectrum. Furthermore, for a Fermi sea one would expect a heat capacity proportional to the temperature. Hence, it should be rather interpreted as a ground state energy that depends on the density. 

Somewhat similarly, the entropy and the $-\sqrt{\lambda}N_c T/2$ term in the chemical potential suggest a ground state degeneracy proportional to the total baryon number, or $\sqrt{\lambda}/2$ times the quark number.
It may be an interesting exercise to write down a distribution and density of states that reproduces this.

The interpretation of the system as being in a ground state is also consistent with the large-mass limit. There, the energy of the system is approximately the baryon number times the baryon mass scale $N_c M_q$, or
precisely the quark number times the quark mass. The cross-over between the high and low density limits occurrs as the induced length scale $\sqrt{\frac{N_f}{\lambda\, \rho}}$ is of the order of the quark mass. 

To obtain the pressure, we have to keep in mind that $\bar{F}$ is densitized. Using $\partial_V G(\rho)|_N = -\frac{\rho}{V} \partial_\rho G(\rho) $ for some function $G$ that only depends on $\rho$, we can obtain $\bar{P} = -\bar{F} + \barho \partial_\barho \bar{F}$:
\begin{equation}
P \ = \  \sqrt{\frac{\lambda}{N_f}}\frac{N_c}{2}\rho^{3/2} \left( {\textstyle \frac{1}{2}}  \Phi(\barho) \, -\, \rho \partial_\rho \Phi(\barho) \right) \, \sim \,
\left\{\begin{array}{ll} 
0 + 0.59 \frac{N_c \lambda^2 \rho^3}{N_f^2 M_q^2}& \ \ : \frac{\rho}{ N_f} \ll \left(\frac{2M_q \!}{\sqrt{\lambda}}\right)^{\! 2} \\
 \frac{1}{6}\mathcal{K}(\frac{1}{\sqrt{2}})\, N_c \rho^{3/2}\sqrt{\frac{\lambda }{ N_f}} & \ \ :
\frac{\rho}{ N_f} \gg \left(\frac{2M_q \!}{\sqrt{\lambda}}\right)^{\! 2}  \ .
\end{array}\right.
%
\end{equation}
From the pressure, we can then obtain the isothermal and adiabatic bulk moduli $\KkappaK_T$ and $\KkappaK_S$. Since the pressure is an intrinsic quantity, we can compute the volume derivative as $\bar{\KkappaK}_T = - V \left. \frac{\partial \bar{P}}{\partial_V}\right|_{T,N} = \barho \partial_\barho\left(-\bar{F} + \barho \partial_\barho \bar{F} \right)=  \barho^2 \partial_\barho^2 \bar{F}$:
\begin{equation}
\KkappaK_T \ = \  
%
%
 \sqrt{\frac{\lambda}{N_f}}\frac{N_c}{2}\rho^{2}
\left(\rho^{3/2} \Phi(\barho)\right)
\, \sim \,
\left\{\begin{array}{ll} 
0 + 1.76 \frac{N_c \lambda^2 \rho^3}{N_f^2 M_q^2}& \ \ : \frac{\rho}{ N_f} \ll \left(\frac{2M_q \!}{\sqrt{\lambda}}\right)^{\! 2} \\
 \frac{1}{4}\mathcal{K}(\frac{1}{\sqrt{2}})\, N_c \rho^{3/2}\sqrt{\frac{\lambda }{ N_f}} & \ \ :
\frac{\rho}{ N_f} \gg \left(\frac{2M_q \!}{\sqrt{\lambda}}\right)^{\! 2}  \ .
\end{array}\right.
%
\end{equation}
and because the full, extrinsic, entropy is just given by the baryon number, we have $\KkappaK_T=\KkappaK_S$. 
Overall, these results suggest that at small densities compared to the mass, we are dealing with some type of pressureless gas of quarks, where the pressure scales proportional to $T^4$ and $\rho^3$, compared to $\rho T$ for an ideal classical gas. As the density increases above the mass scale, the zero-temperature pressure and bulk modulus become large. This is consistent with the density dependence of the ground state energy and with a length scale induced by the finite density.

We can also obtain the speed of sound. For the small-density limit this is just the non-relativistic ${v}_s^2  = \frac{\KkappaK}{M_q N_c \rho} $, which vanishes because the bulk modulus vanishes. In general we can use the fact that the energy density depends only on the density and obtain $v_s^2 = \left.\frac{\partial P}{\partial U}\right|_S = \left(\frac{\partial U}{\partial \rho}\right)^{-1} \frac{\partial P}{\partial \rho} = \frac{\KkappaK}{\rho \partial_\rho U}$:
\begin{equation}
v_s^2 \, = \,  
\rho \partial_\rho \ln \left( \partial_\rho 
\left(\rho^{3/2} \Phi(\barho)\right) \right) 
\, \sim \,
\left\{\begin{array}{ll} 
0 + 1.76 \frac{N_c \lambda^2 \rho^3}{N_f^2 M_q^2}& \ \ : \frac{\rho}{ N_f} \ll \left(\frac{2M_q \!}{\sqrt{\lambda}}\right)^{\! 2} \\
\frac{1}{2} & \ \ :
\frac{\rho}{ N_f} \gg \left(\frac{2M_q \!}{\sqrt{\lambda}}\right)^{\! 2}  \ .
\end{array}\right.
%
\end{equation}
The density-independent result in the large-density limit is dictated by causality.
It would be very interesting to compare these results with the ones obtained from the normal modes in the gravity side as in \cite{zerosound,luttinger}.
\subsection{Subleading Terms}
To study the subleading terms, we obtain the difference between the low-temperature limit for $F$ from equation \reef{keyeq} and the full numerical result that we can obtain from the methods in \cite{fancytherm}.

To do so, we proceed as follows:
We fix $\Psi_0$ and then vary $\tilde{\rho}$. Beyond the dominant constant term, there is some small temperature dependence in $\barho$
that appears in the numerical results at $\order(\baT^3)$.
This is compensated for by using the actual numerical value $\barho(\Psi_0,\baT)$
in \reef{fformula} and compare this to the numeric result for the same $\Psi_0$ and $\baT$. From that point on, we ignore the small variation in $\barho$ and consider it fixed. It turns out to be most convenient to consider a temperature range over which the significant subleading term in $\bar{F}$ is suppressed by at least $10^{-4}$ and $\barho$ varies at most by a factor of $10^{-3}$. Because the numerical accuracy in $\bar{F}$ is approximately $10^{-7}$, the systematic errors from ignoring the variation in $\barho$ appear then roughly at the same level as the numerical noise. 

\begin{figure}
\includegraphics[width = 0.49\textwidth]{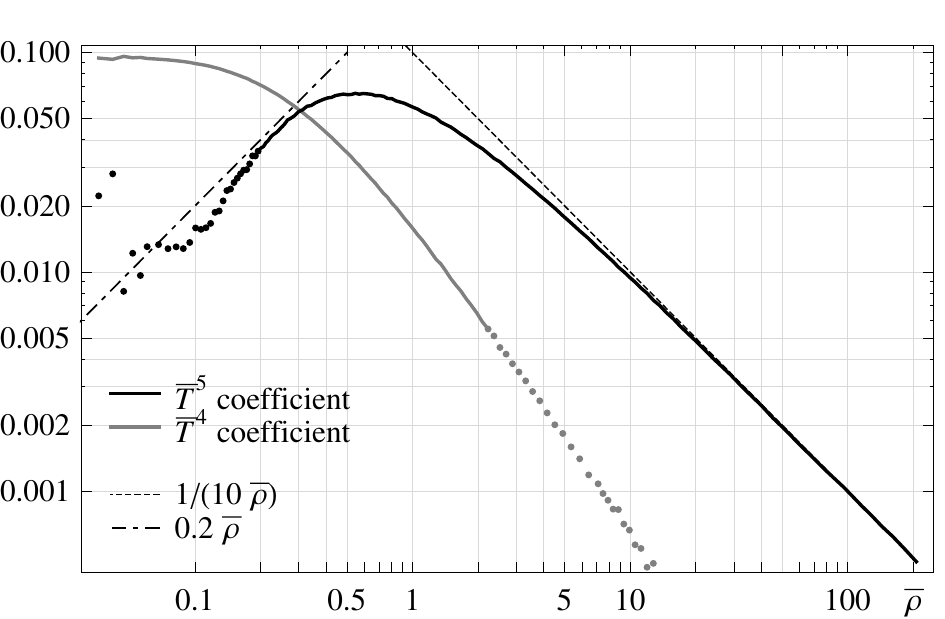}\hspace{0.02\textwidth} \includegraphics[width = 0.49\textwidth]{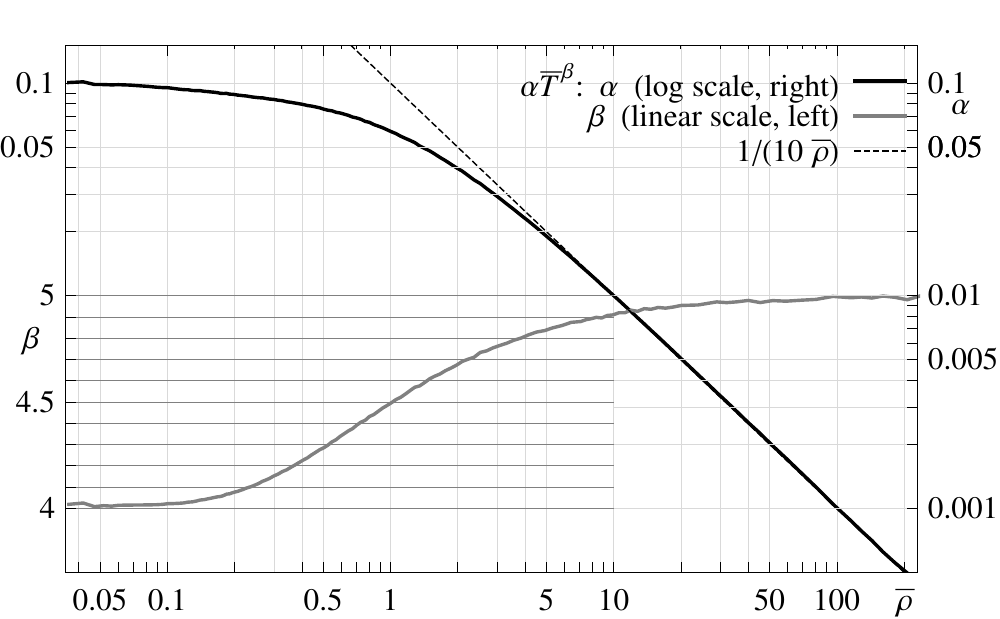} 
\caption{The coefficients in the subleading term in the free energy parametrized as $\delta \bar{F} = - a \baT^4 - b \baT^5$ (left) and $\delta \bar{F} = - \alpha \baT^\beta$ (right). On the left, the dots indicate a regime in which the fits for the smaller term are not numerically reliable anymore.}\label{tempexpplot}
\end{figure}

It turns out that this subleading term is always negative and interpolates from approximately $\delta \bar{F} \approx - \frac{\baT^4}{10}$, i.e. $\bar{c}_v \approx \frac{6\baT^3}{5}$, at small density to precisely $\delta \bar{F} = - \frac{T^5}{10 \barho}$ at large density. The latter is consistent with the results in the massless case in \cite{fancytherm} and \cite{zerosound}, from which we expect at large densities $\bar{c}_V = 2 \frac{T^4}{\barho}$.
The most convenient ways to parametrize the numeric results are $\delta \bar{F} = - a \baT^4 - b \baT^5$ and $\delta \bar{F} = - \alpha \baT^\beta$, which are not distinguishable at the level of the straightforwardly achievable numerical accuracy and are shown in figure \ref{tempexpplot}.
An exponential suppression of the subleading term, that would be suggestive of an energy gap, is however ruled out.
\section{General cases}\label{gencas}
Even though it is not the focus of this paper, let us briefly comment about what happens in other cases than the 2+1 dimensional defect. In principle, the methods discussed in sections \ref{limit} and \ref{thermo} are straightforwardly generalized to fundamental matter supported in (d+1) dimensions. In the following $\tilde{(\cdot)}_d$ and  $\bar{(\cdot)}_d$ will denote dimensionless quantities analogous to the definitions in sections \ref{setup} and \ref{limit}. 
The setups that come most straightforwardly to  mind are the usual supersymmetric D3-D7 and D3-D3 setups that are T-dual to the D3-D5 case,
\begin{equation}
\begin{array}{rccccc|c|cccccl}
  & & 0 & 1 & 2 & 3 & 4& 5 & 6 & 7 & 8 & 9 &\\
  & & t & x & y & z & r&   &   &   &   &  \theta &\\
\mathrm{background\,:}& D3 & \times & \times & \times & \times & & &  & & & & \\
\mathrm{probe\ (1+1)\,:}& D3 & \times & \times &  &  & \times  & \times &  & &  & &  \ \ \mathrm{or} \\
\mathrm{probe\ (3+1)\,:}& D7 & \times & \times & \times & \times & \times  & \times & \times & \times &  & &  \ \ \ 
\end{array} 
\labell{arraymany}
\end{equation}
with the induced metric and the action given analogous to equation \reef{branemetric} and \reef{braneaction}, respectively. For the action in the form of equation \reef{backact}, we have to keep in mind that in a D3-Dp embedding, the factor $(1-\Psi^2)$ is replaced by $(1-\Psi^2)^{(p-d-1)/2}$, which is in our supersymmetric cases $(1-\Psi^2)^{d/2}$.
The solutions for the gauge field are
\begin{equation}\label{shita}
\partial_u \tilde{A}_{d\,t}(u) \ = \ \frac{\tlrho\sqrt{1 + u^2 h(u) \frac{\Psi'(u)^2}{1- \Psi^2(u)}}}{\sqrt{  \tlrho_d^2 u^{2d} + (1-\Psi(u)^2)^{(p-d-1)}}}
\end{equation}
but we will not review further details of the backgrounds as they are readily found in the literature \cite{winters,robdens,luttinger} and they are not essential here.
The most important property that we need to know for the the embeddings $\Psi$ at this point is that near $u\rightarrow 0$, in the D3-D7 system $\Psi\sim \tlm_3 u + \tlc_3 u^3$ and $\Psi\sim \tlm_1 u + \tlc_1 u\, \ln\, u$ in the D3-D3 system. 

\begin{figure}
\includegraphics[width = 0.49\textwidth]{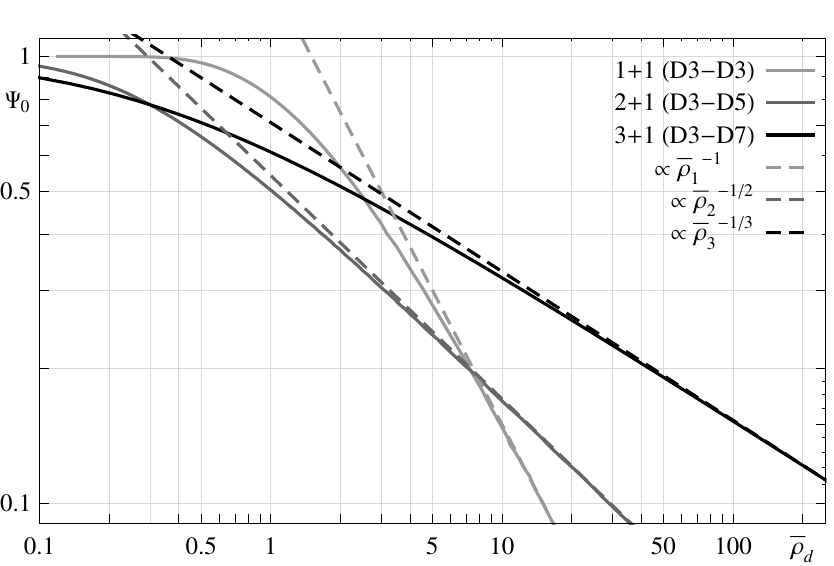} 
\hspace{0.02\textwidth} 
\includegraphics[width = 0.49\textwidth]{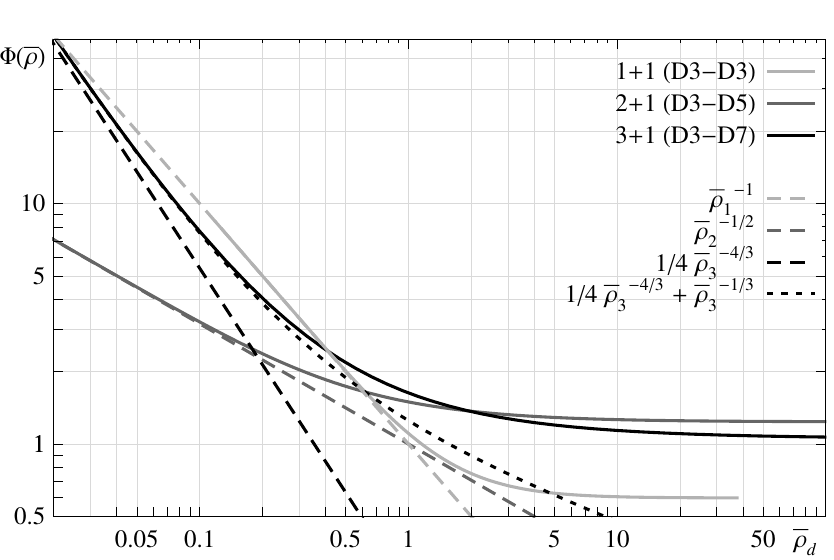} 
\caption{Left: The boundary value $\Psi_0$ as a function of the density $\barho$ for various setups. The large-density approximations are shown dashed. Right: $\Phi_d$ as a function of the density $\barho_d$, illustrating also the $\barho \ll 1$ limit.}\label{lastgoddamnplot}
\end{figure}

To obtain the brane profiles, the radial coordinate has to be defined by $\xi_d = \frac{u}{\tilde{\rho}_d^{1/d}}$, and the large-$\xi_d$ expansions of $\Psi$ are $\Psi_0 - \frac{1}{14 \xi_3^6} \Psi_0(1-\Psi_0^2)^3$ and $\Psi_0 - \frac{1}{6 \xi_1^2} \Psi_0(1-\Psi_0^2)$. We show the corresponding $\Psi_0$ as a function of the density $\barho_d = \left. \partial_{\xi_d} \Psi\right|_{\xi_d \rightarrow 0}$ in figure \ref{lastgoddamnplot}.

We find that at large densities, $\Psi_0$ scales proportional to  $\barho_d^{-1/d}$. As in the 2+1 case, there seems to be an exact relation $\bac_3 = \barho_3 \frac{\Psi_0}{ 2}$ and $\bac_1 = \barho_1 \Psi_0$.

To compute the thermodynamics, we first note that the overall factor $\barho^{3/2}$ in the free energy in equation \reef{keyeq} becomes $\barho^{(d+1)/d}$. 
In analogy to \reef{keyeq}, the factor in the integral coming from the AdS structure is $\int\! \frac{d\xi_d}{\xi_d^{d+2}}$, and in the large-$\xi$ limit, the solution \reef{shita} gives us a term $\xi_d^2$ in the Legendre-transformed action such that we can reproduce the $\xi \in ]\sqrt{\tlrho},\infty[$ part of the integral \reef{keyeq}. Hence, the free energy becomes
\begin{equation}
\bar{F}_d \ = \ \barho_d^{(d+1)/d} \Phi_d(\barho_d) \, - \, \barho_d \baT \ + \ \order(\tlrho^{(1-d)/d}) \ 
\end{equation}
and the entropy is just the baryon number as in the 2+1 dimensional case.

\subsection{3+1 Dimensions}
To discuss $\Phi_d(\barho_d)$ let us start with the 3+1 dimensional system. In this case, the counter terms from \cite{skenderis} are structurally the same as in the D3-D5 case, and hence the integral is similar to \reef{keyint}, up to the appropriate powers in the expression. The result is shown in figure \ref{lastgoddamnplot} in comparison to the 2+1 defect.

We find that the power scaling at small densities approaches $\Phi_3(\barho_3) \sim \frac{1}{4 \barho_3^{4/3}}$. This means that the leading behavior in $\barho_3$ is $\bar{F}_3 = const. - \barho_3 \baT_3$, i.e. the free energy (and also the internal energy) per unit volume (given by the meson mass scale) is proportional to the mass -- independent of the baryon number density. Essentially, this implies that the configuration is unstable in this regime, with a negative pressure proportional to $-M_{meson}^4$. As the chemical potential is monotonously increasing (i.e. we have a positive ``density of states'') this instability is different from the one studied in \cite{yunseok}. Furthermore, there the ``funnel'' on the gravity side is becoming very narrow, whereas here it has narrow but finite width and is becoming very long. Still, $\Psi'$ is becoming very small and hence the tension of the brane that falls into the horizon is close to the tension of an appropriate number of fundamental strings attached to a brane that does not extend down to the horizon.

The subleading term in $\bar{F}_3$ is just $\barho_3^{-1/3}$, giving again just the baryon number times the meson mass scale as we observed in the 2+1 system. 

At large densities, $\Phi_3$ becomes constant, close to 1, so the free energy density is approximately $\bar{F} \sim \barho (\rho^{1/3} - \baT)$ which is consistent with the picture of an induced length scale discussed in section \ref{thermquan} and mimics again a Fermi surface. The sound speed in this limit is now $c_V^2 = \frac{1}{3}$ as in \cite{isospin,responses}.
\subsection{1+1 Dimensions}
The D3-D3 system is profoundly different. While the overall prefactor of the integral in \reef{keyint} and first counter term in \reef{ibdy} just follow dimensional arguments, there are  a few logarithmic divergences. Hence, the appropriate counter term from \cite{skenderis} for the scalar is now $\frac{1}{2}\sqrt{\gamma}\Psi(\epsilon)^2(1+1/\ln(\epsilon))$. 
In contrast to the 3+1 and 2+1 dimensional systems, where the boundary term gave us the appropriate variation of the action w.r.t. the scalar, $\tlc\,\delta \tlm$, the variation is now $-\tlm\, \delta \tlc$. Hence, we have to do a ``Legendre transformation'' and add the extra boundary term $\tlc\, \tlm$ to the action in order to consider a fixed mass.
This counter term cures the divergence coming from the scalar but there remains one divergence 
coming from the logarithmically divergent $A_t \sim \tlmu +  \ln \epsilon$. 
While the variation of the action is finite, and also the Legendre transformation can be trivially made finite by using $\tlrho (A_t - \tlrho\ln\epsilon)$, the action itself is divergent. 

In principle, one can add extra boundary terms to cure this divergence.
For example in \cite{luttinger}, the authors use a counter term $\frac{A_\mu A_\nu \gamma^{\mu\nu} \sqrt{\gamma}}{2\, \ln\, \epsilon}$ which has, however, a non-trivial variation $\frac{A_\mu A_\nu \gamma^{\mu\nu} \sqrt{\gamma}}{ \ln\, \epsilon}\delta  A_\nu$.
In the limit $\epsilon \rightarrow 0$ and for a finite variation $\delta A_t$, this counter term is identical to the Hawking-Ross boundary term \cite{hawkingross}, and cancels precisely the boundary term in the variation of the gauge field. This is because, just as with the scalar, implicitly this term also includes a Legendre transformation, and in terms of the electric flux (or density) the asymptotic variation of the gauge field is $\delta A_t \sim \tlrho \ln \epsilon$, so it diverges at the boundary and results in a finite variation $\mu \, \delta \tlrho$. Coincidentally, this counter term is identical to adding $\frac{\tlrho^2}{2} \ln\epsilon$ and then doing the finite Legendre transformation -- or first doing a (diverging) Legendre transformation $\tlrho A_t$ and then adding $-\frac{\tlrho^2}{2} \ln\epsilon$.

For practical aspects, the integral for $\Phi$ does converge only logarithmically in the UV limit, so we have to take care of this using approximate analytical solutions to extend the numerical result to even smaller $x$.
We show  the result for $\Phi_1$ in figure \ref{lastgoddamnplot}. There, we see that $\Phi$ converges to approximately $0.596$ at large densities $\barho_1 \gg 1$, giving us physics with an induced length scale similar to the other cases, and now a speed of sound of $v_s^2 = 1$, which is again the ``conformal'' result. 
At small densities or large masses $\barho_1 \ll 1$, $\Phi_1$ approaches $\Phi_1\sim \frac{1}{\barho_1}$, which implies again that the energy of the system is described by a pressureless gas of mesons. Similarly, the subleading term giving rise to the pressure implies that that the pressure is at least of order $\barho^4$, but numerical accuracy implies that we cannot give a more detailed result and also cannot exclude a very small but finite (positive or negative) pressure at vanishing density.

One might worry in how far the redefinition $u\, \tlrho_1 = \xi_1$ in the logarithmic terms causes some mixing between the density and the temperature. As the final action however is free of logarithmic divergences, there should be no such remaining terms. Furthermore, we see that our results for the free energy rely certainly heavily on using the correct boundary terms. Further discussions of peculiarities of this 1+1 dimensional system can be found in \cite{luttinger}, which focuses on this case, albeit not in this limit.
\section{Conclusions}\label{conclu}
In this paper, we used holography to explore the low-temperature regime of fundamental matter with finite mass and baryon number density coupled to the usual $N_c \gg 1$, $\mathcal{N}=4$ SYM theory above the deconfinement phase transition. The fact that we used a top-down approach guarantees us that the field theory is consistent and well defined, giving more relevance to our findings. Using $N_f$ Dp probe-branes in a background of $N_c \gg N_f$ D3 branes, we found a new temperature-independent scaling solution of the probe brane embeddings. On the field theory side, this manifests in the fact that the low-temperature physics is only governed by the density to mass ratio, $\rho / M_q^2$ and interpolates between a mass-dominated low-density regime and a density-dominated large-density regime. On the gravity side the radius-dependent size of the compact sphere of the probe brane geometry forms a long ``funnel'' of finite size, with the scaling solution in the asymptotic ``opening''.

First, we concentrate on a defect supported on 2+1 dimensions that is dual to a D3-D5 setup.
From the usual thermodynamic relations, we obtained the properties of this matter, which suggest that there is a new kind of quantum liquid. The entropy is just given by the baryon number times $N_c \sqrt{\lambda}/2$ or the quark number times $ \sqrt{\lambda}/2$, which implies a large ground state degeneracy. Consistent with this, the chemical potential has a negative term $-T N_c \sqrt{\lambda}/2$, that is consistent with a Bose-Einstein distribution with a degeneracy proportional to the baryon number.

%
%
%
%
%
%
%
The heat capacity however vanishes completely in the scaling solution, which means that it is of higher order in temperature than a classical Fermi or Bose gas. To explore the possibility of a gap, we explored the subleading terms in the free energy numerically. It turns out that the heat capacity is not exponentially suppressed -- so there is no energy gap -- but rather interpolates between $c_V\propto T^3$ at small densities and $c_V \propto T^4$ at large densities; in the latter case  the coefficient matches with the analytic results of the massless case in \cite{zerosound} and \cite{fancycon}.

In the small-density limit, the internal energy is just given by the baryon number times the baryon mass scale, or precisely the overall quark number times the quark mass.
At the same time, the pressure, bulk modulus and sound speed vanish; suggesting that the system becomes a pressureless gas of quarks. This is, however, still not a classical non-interacting gas, as the leading terms in the pressure are of higher order in temperature and pressure than for a classical gas, i.e. they vanish more quickly as $T\rightarrow 0$ and $\rho \rightarrow 0$.

%
%
%
%
%
At large densities, the internal energy is given by the quark number times an energy scale that is proportional to $\sqrt{\rho}$ with a corresponding leading term in the chemical potential proportional to $\sqrt{\rho}$. This is, however, not to be interpreted as a Fermi surface, but rather as a length scale of the order of $\sqrt{\frac{ N_f}{\lambda\, \rho}}$
that is induced by the finite density at strong coupling and determines the ground state energy. Note that this is much shorter than just the geometrical separation of quarks. Consistent with this, the pressure and bulk modulus are proportional to  $N_c \rho \sqrt{\frac{\lambda\, \rho}{ N_f}}$
and the sound speed becomes $\frac{1}{\sqrt{2}}$ times the speed of light.
The cross-over between the high and low density regimes occurrs when the induced length scale is of the order of the quark mass. 

We also looked at the cases of fundamental matter supported in 3+1 (D3-D7) and 1+1 (D3-D3) dimensions. The embeddings and the general structure of the thermodynamics are very similar, but there are also some important differences. In the 3+1 case, the system exhibits a new thermodynamic instability  characterized by a negative pressure at small density, and in the 1+1 case it turns out that there are some non-trivial logarithmic divergences. Applying our analysis to this case and adding additional 
counter terms, such as the one discussed in \cite{luttinger}, yields qualitatively similar results to the 2+1 dimensional case.

To study the properties of this quantum liquid a bit further -- in particular in the 2+1 defect case -- it would be interesting to obtain properties that are obtained from linear response theory and two point functions, such as the viscosity, thermal and electric conductivity, quasiparticle spectrum and also the speed of sound to confirm consistency of the results.
On the gravity side, these are straightforwardly obtained from the equations 
of motion for perturbations of the world-volume gravitational and gauge fields. Certainly, there may be some mixing between the different modes, but dealing with coupled systems in probe-brane backgrounds both analytically and numerically is straightforward \cite{fancycon,luttinger,recenttwist}. Furthermore, the equations should simplify by following the limiting procedure  considered in section \ref{limit}.

In the 3+1 dimensional case, it would be interesting to further explore this new instability, and it would be also interesting in both the 3+1 and 1+1 cases to study the physics and subleading terms more thoroughly and to study dimensional peculiarities in the 1+1 case.

Certainly, it would also be interesting to see in how far some of the properties that we find are realized in experimentally studied systems.

\acknowledgments The author would like to thank Bum-Hoon Lee for useful comments on the draft of this paper, Janet Hung and Aninda Sinha for helpful clarifications regarding \cite{luttinger} and
Sang-Jin Sin, Yun-Seok Seo and Yumi Ko
for useful discussions.
This work was supported by the National Research Foundation of Korea(NRF) grant funded by the Korea government(MEST)
through the Center for Quantum Spacetime(CQUeST) of Sogang University with grant number 2005-0049409.
%
%
%
%
%
%
\bibliography{mybibfile}
\bibliographystyle{JHEP}
\end{document}